# Differential Music:

# Automated Music Generation Using LSTM Networks with Representation Based on Melodic and Harmonic Intervals

**Hooman Rafraf**

University of Florida

hooman.rafraf@gmail.com

## Abstract

This paper presents a generative AI model for automated music composition with LSTM networks that takes a novel approach at encoding musical information which is based on movement in music rather than absolute pitch. Melodies are encoded as a series of intervals rather than a series of pitches, and chords are encoded as the set of intervals that each chord note makes with the melody at each timestep. Experimental results show promise as they sound musical and tonal. There are also weaknesses to this method, mainly excessive modulations in the compositions, but that is expected from the nature of the encoding. This issue is discussed later in the paper and is a potential topic for future work.

## 1  Introduction

Music generation using artificial intelligence with artistic style imitation has been broadly explored within the past three decades. A critical step in this task, whether in audio domain or symbolic domain, is that of converting music into machine-perceivable matrices of numbers. In the symbolic domain which is the focus area here, the MIDI format provides a powerful tool to facilitate such a conversion. Every standard musical pitch in MIDI format is assigned a unique number which can be extracted using parsing tools and written to matrices. The numbers must be further processed before being ready for use as inputs to AI models. This pre-processing of data is gravely impactful on the performance of the model and therefore must be carefully thought through. It is also an area of creativity and exploration since it can be done in a great multitude of ways. The result of this preparation step which can be considered as an encoding of the original data is commonly referred to as "representation" in the machine-learning terminology. The focus here is on the representation of music specifically for use with LSTM-based models. Music encoding is done using a novel method that is based on musical intervals rather than individual pitches. I will train a model and generate results using the new encoding system and will discuss the benefits and flaws of it compared to the existing representations.

## 2  Background

### 2.1  Recurrent Models

Using recurrent neural networks (RNNs) for music composition is at least as old as the research by Todd (1989) where he trained a small RNN with 15 hidden units to compose monophonic music. These networks suffer from a problem known as "vanishing gradients" which prevents them from learning long-term dependencies (Hochreiter et al. 2001). LSTM is a special type of recurrent network that performs better in learning temporal dependencies (Hochreiter and Schmidhuber 1997) which makes them more suitable for music composition than the original RNN. The LSTM can learn long-term structure in music, resulting in better formal coherence in the composition. This was shown by Eck and Schmidhuber (2002) for the first time by training an LSTM model to compose blues music. Since then, many have designed and implemented successful music generation models using recurrent neural networks. Transformers (Vaswani et al. 2017) are another type of recurrent networks that vastly outperform LSTMs at learning long-term dependencies in sequential data. OpenAI's MuseNet (Payne 2019) which uses an optimized version of Transformers called Sparse Transformers (Child et al. 2019), and Music Transformer (Huang et al. 2018) developed at Google under the Magenta Project both accomplished state-of-the-art music generation using Transformer-based models. I use LSTM because of its better accessibility but am positive that the encoding proposed in this paper can improve results of Transformer-based models as well.

### 2.2  Existing Representations

To properly represent music for machine-learning tasks one must account for the following: At what moment is the onset of each note in the music, and how long each note lasts. Though some authors such as Oore et al. (2020), Zhao et al. (2019), Payne (2019) and Huang et al. (2018) have also included dynamics and/or rubato information in the representation to give the results human expressivity, that is not the focus in this paper. A common approach to address the mentioned questions is to quantize time into small timesteps and allocate a pitch vector for each timestep that contains information on the active notes in that timestep. This vector commonly has one node per note in the pitch range of the music that is either *1* indicating that the note is sounding in that particular timestep, or *0* indicating otherwise. In machine-learning jargon a vector where one node is *1* and the rest are *0* is called a "one-hot" vector. In case there are multiple *1*'s the vector can be called "multi-hot". The current encoding is an ordered collection of one-hot or multi-hot vectors (depending on the number of voices) and can be seen as a matrix view of the piano-roll representation, as seen in Figure 1. It has proven successful in works by Oore et al. (2020), Kumar and Ravindran (2019), Zhao et al. (2019), Eck and Lapamle (2008), and Boulanger-Lewandowski, Bengio, and Vincent (2012) among others.

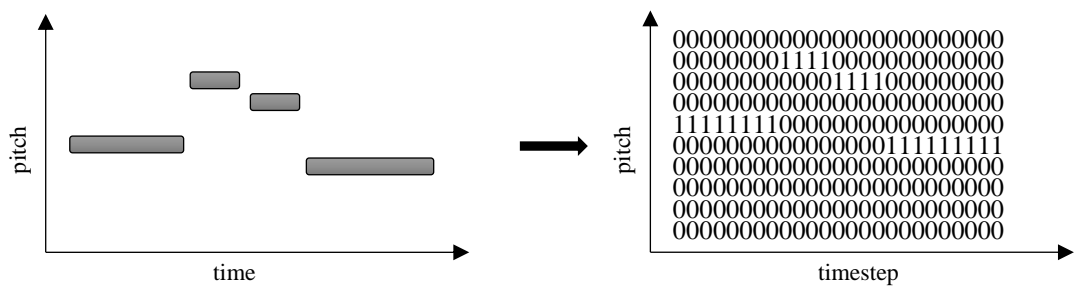

*Figure 1.* Conversion of piano-roll to matrix representation using one-hot vectors for monophonic music. For polyphonic music, the vectors will be multi-hot with a 1 for each active note in the timestep.

Another popular approach is text-based representation where notes are indicated by a sequence of characters typically separated by a delimiter, e.g., "C4|D4|B3". BachBot (Liang et al. 2017), DeepBach (Hadjeres, Pachet, and Nielsen 2017) and work by Choi, Fazekas, and Sandler (2016) are a few examples. "ABC notation" is another text-based convention as seen in work by Sturm et al. (2016).

In terms of rhythm, the problem with quantizing time into equal slices is that re-articulation of the same note in a row becomes ambiguous. For instance, four successive sixteenth notes of the same pitch, say A5, are encoded the same as one quarter-note at A5. Assuming time is quantized into sixteenth-note timesteps, in both cases the representation is four vectors with a *1* at the index corresponding to A5 and zeros everywhere else. This is one reason why some authors such as Kumar and Ravindran (2019) and Liang et al. (2017) prefer to handle rhythm differently and encode note durations in a separate vector or as appendix of the pitch vector. I choose to employ the time-quantization approach since the chorale music used for training seldom contains repeated notes, therefore the benefit of simplicity in the representation outweighs the mentioned disadvantage.

## 3   Absolute vs. Relative Pitch

### 3.1   Problem

Even though transposing the entirety of a piece could affect our listening experience and how we perceive the music, it does not affect our recognition of it. Two pieces that are transpositions of each other will be "recognized" as the same piece but in different keys or registers. In that regard the matrix representation of piano-roll described earlier is not efficient. If similar musical patterns appear in different keys in the training set, this representation encodes them differently because the patterns are formed by different sets of pitches. As a result the model fails to learn from that pattern unless it appears enough times in various keys in the training set.

There are two common ways around this problem within the existing works. Some authors transpose all the pieces in the training set to the same key. We see this in works by Choi, Fazekas, and Sandler (2016), Liang et al. (2017), Mozer and Soukup (1990), Boulanger-Lewandowski, Bengio, and Vincent (2012), Eck and Lapamle (2008), and Kumar and Ravindran (2019). This will improve the correlation between musical similarity and encoding similarity, thus helping the model learn the temporal patterns better. But this improvement is very limited since even in the same piece of music similar patterns can appear in different octaves or in different keys, e.g., in case there is tonicization or modulation in the piece. Even in the same key and octave, a melodic pattern can appear several times in a row, each time from a different starting point. This structure is known as "sequence" in music and is a common compositional technique (Benward and Saker 2003).

Another workaround is to augment the data by transposing the pieces to various keys and include them in the training set. Authors who chose this approach include Oore et al. (2020), Hadjeres, Pachet, and Nielsen (2017), and Eck and Lapamle (2008). This method will mitigate the problem to some extent by introducing the model to musical patterns in different keys, improving the chance for the model to see similar patterns in the same key that would otherwise appear only in different keys. But this approach comes at the cost of enlarging the training set enormously. It is typical that only a handful of transpositions are added since it is quite impractical to add a version of each piece in all possible keys in every octave.

## 3.2 Solution

Using relative pitch instead of absolute pitch solves the problem entirely, which is the central proposition of this paper. In monophonic music, given the initial note, any melodic line can be seen as a sequence of intervals rather than a sequence of pitches. For instance, a melodic sequence of notes designated by the MIDI pitches [67, 70, 69, 67] would be denoted by [3, -1, -2], and later for reconstruction the cumulative sums of the numbers from the latter vector would be appended to [67], or to any other desired opening note. The choice of the opening note only specifies the key and register in which we would like to hear the music. We lose this information with this encoding, which is nothing unfortunate, quite on the contrary. Now all the melodic patterns that are similar or identical would be encoded similarly or identically. Our vector [67, 70, 69, 67] denotes the same melody as [40, 43, 42, 40] though in a different key and register, but it is comprised of totally different numbers. But in the new encoding both vectors are represented by [3, -1, -2]. This is advantageous in many ways. The training data is more informative for the model since patterns in the training data are more "visible". There is no more need for the data to be transposed to the same key or be augmented, and even though the training set is smaller the learning potential for the model is higher. The proposed encoding can be viewed as a lossless compression by a large factor. Imagine how many different vectors obtained using the original encoding would all be encoded as [3, -1, -2] in the new encoding. This tremendously reduces the size of the training set yet maintains the same amount of information.

It is common approach to convert the sparse vectors to one-hot vectors as the model seems to learn significantly better (see "Representation" section for implementation details). By employing the new encoding, after the conversion the dimensionality of the input space will also be smaller compared to the original encoding, which adds to the factor of data compression. The length of the encoding vector for melodic lines would not need to exceed the largest leap in the melody which is almost never greater than one octave, whereas in the original encoding the length must at least span the range of pitches from the lowest note to the highest note in all the training data.

## 3.3 Polyphony

To progress from monophonic to polyphonic music where multiple notes can sound simultaneously, we can apply the same idea to chords. Since the chord quality does not change with transposition, we can encode chords by specifying the interval that each chord note makes with the root or any other desired note. Similar to what was discussed for melodic lines, the choice of the origin note does not change the quality or spacing of the chord, only the height of the sound. With the original encoding it is less complex to allow polyphony. It only suffices to use multi-hot vectors instead of one-hot vectors, accounting for all the notes that sound simultaneously at each timestep. But in the new encoding the melody and chord vectors are obtained separately and must be concatenated. See the next section for details.

## 4   Representation

A common way of measuring intervals in music is to count the number of half-steps that the interval contains. I make two vectors for each timestep of the music. The first is a one-hot vector that encodes melodic motion of the soprano by specifying the number of half-steps it moved from the previous timestep. The second is a multi-hot vector that encodes the rest of the active notes in the timestep in terms of their relative distance, in half-steps, to the soprano. Let's call the former *melody vector* and the latter *chord vector*.

In melody vector the index of the active node indicates the number of half-steps in the melodic interval. I allow a maximum of 11 half-steps (just short of one octave) for leaps in melody in each direction. Any melodic interval larger than 11 half-steps will be brought within the octave using a mod-12 operation. We also need one node whose activation indicates interval of zero, meaning the melody has not moved from the previous timestep. This sums the size of the melody vector to 11+11+1=23. A quick analysis of the dataset showed that among the total of 22,688 melodic motions of soprano in all the chorales only 51 are octave leaps or larger. Modifying such a small fraction of training data is not detrimental to the learning process. This makes the mod-12 operation a logical choice since it greatly reduces the size of the input vectors for melody, which consequently reduces training time and increases the performance of the model. Figure 2 illustrates the structure of the melody vector. Pauses in the music are not encoded; all silent timesteps are omitted prior to encoding. Incorporating pauses will be a potential addition of future work.

| index | 1 | 2 | 3 | 4 | 5 | 6 | 7 | 8 | 9 | 10 | 11 | 12 | 13 | 14 | 15 | 16 | 17 | 18 | 19 | 20 | 21 | 22 | 23 |
|-------|---|---|---|---|---|---|---|---|---|----|----|----|----|----|----|----|----|----|----|----|----|----|----|
| interval | 11 | 10 | 9 | 8 | 7 | 6 | 5 | 4 | 3 | 2 | 1 | 0 | 1 | 2 | 3 | 4 | 5 | 6 | 7 | 8 | 9 | 10 | 11 |

down                                                up

*Figure 2.* The melody one-hot vector has the length of 23. It can specify melodic motion of up to 11 half-steps up or down. Activation of the node in the center indicates that the previous note has sustained.

In chord vector the index of each active node represents the number of half-steps the corresponding chord note makes against the soprano. The length of this vector (denoted *maxSpace*) is determined by the largest distance between the soprano and the bass note in all the training data. If the melody is unaccompanied at a timestep, the chord vector corresponding to that timestep will contain all zeros. Figure 3 shows an example of the chord vector corresponding to a second-inversion B-diminished triad. The D in the soprano is accounted for in the melody vector. The chord vector has two active nodes at indices 3 and 9, indicating the harmonic intervals between the two bottom notes and the soprano.

Finally, the two vectors are concatenated into one with size 23+*maxSpace* which will be a single timestep for the input of the LSTM.

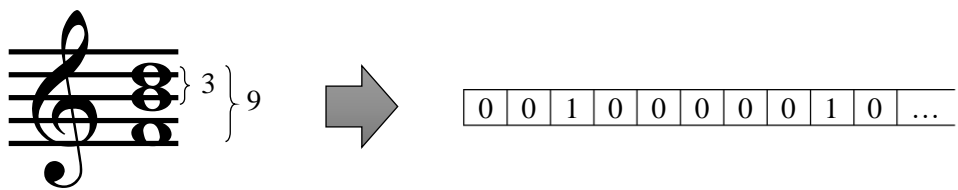

*Figure 3.* Obtaining the chord vector that corresponds to three notes sounding simultaneously. The length is determined by the largest harmonic interval in the training data.

## 5  Data Preparation

The set of training examples used for this experiment was "JSB Chorales" (Boulanger-Lewandowski, Bengio, and Vincent 2012) which contains 382 chorales composed by Bach and is available for download in MIDI format from Papers With Code at the following URL: https://paperswithcode.com/dataset/jsb-chorales.

The Matlab library MIDI Toolbox (Toiviainen and Eerola 2017) was used for the initial conversion of the MIDI files to Matlab arrays and from there to Python for encoding. *maxSpace* turned out to be 64 so the length of the encoded vectors is 64+23=87. The dataset had to be corrected first since the temporal positioning of musical beats in the dataset are not aligned with the MIDI beats. But because the error is consistent, the fix was straightforward.

The choice of time quantization in this experiment is that each timestep of the encoded data represents an eighth-note in the music. The optimal quantization could vary among different datasets, but since the chorales rarely contain shorter than eighth notes, this quantization works best for our purpose. I used the first 350 chorales for training and reserved the rest to be used as "prompts" for generation step (see the section Model Training and Prediction for generation details).

The melody vector for the first timestep of each piece needs to be tackled since the soprano in the first timestep does not have a previous note to make a melodic interval with. We can either assume that this note is sustained from an imaginary previous timestep and zero out all the nodes of the vector, or we can entirely skip the first timestep of each piece. Neither approach would make a considerable impact on the training because among a total of 307,960 timesteps in the training examples there are only 350 that mark the beginning of a chorale. I took the second option for ease of coding.

A moving window is then used to slice up the encoded matrices into fixed-length training examples. The size of the window must be chosen carefully. If it is too long the number of training examples might become too small since many of the chorales are shorter than the window size and will be skipped. If the window size is too short the LSTM will not learn long-term structure in the music. In this work after some experimentation the window size was chosen to be 40 which resulted in a total of 7,699 training examples.

## 6  The Experiment

### 6.1  Model Architecture

The function of the model is to receive 40 successive timesteps of encoded music as input and generate a single timestep as output. The output is the prediction of the model as to what the next timestep would have been had Bach written it. If the model is trained properly and working as intended, the prediction should be "correct" with respect to stylistic characteristics of chorale music. The input goes to a two-layer stacked LSTM with 300 and 200 hidden units and *tanh* activation, which is connected sequentially to a dense layer with 200 units and from there to another dense layer with 100 units. Both dense layers have *ReLU* activation. The output is connected separately to a dense layer with 23 units and another dense layer with *maxSpace*(=64) units. These two will generate the melody and chord vector outputs of the network respectively. The former has *Softmax* activation since only one node is active in the melody vector. The latter has *Sigmoid* activation which predicts the chord notes in terms of their distance from the soprano. The model's output is simply the concatenation of the outputs from the last two dense layers. The input shape of the network is {*batchSize* × *timesteps* × *features*}, which in our case becomes {7,699 × 40 × 87(=23+64)}. The model is compiled with *Adam* optimizer and *binary cross-entropy* as loss function. Figure 4 demonstrates the model structure.

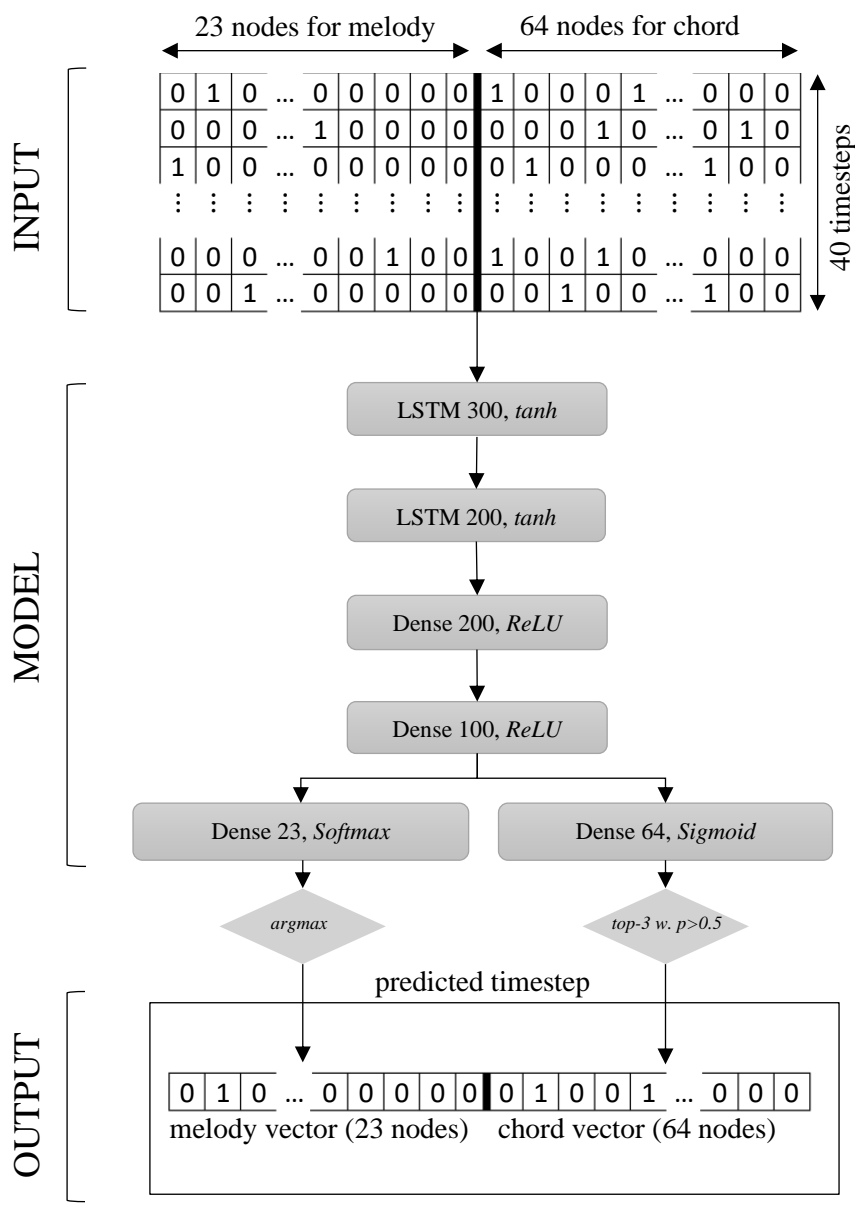

*Figure 4.* Model architecture. The input is a musical prompt with 40 timesteps. The output is a single timestep generated as the prediction of the next timestep. For melody output, the node with highest value (argmax) is activated. For chord output, the three highest nodes are activated on the condition that their value is > 0.5.

The output of the model must be converted from a probability distribution to a vector of zeros and ones. In the melody vector the index with the maximum probability (argmax) is set to 1 and all other nodes are set to 0. In the chord vector the top three nodes with probability > 0.5 are set to 1 and all other nodes to 0. It is best to pick three nodes in the chord vector because in Bach chorales there are predominantly four notes sounding together; the soprano is predicted by the melody vector which leaves the remaining three voices. It is also possible that the output has fewer than three nodes with $p > 0.5$, which means the predicted timestep has fewer than four voices sounding simultaneously. This is okay and is consistent with chorale music.

## 6.2 Model Training and Prediction

Various experiments showed that for the model to generate musically acceptable output, the training loss must reach below 0.001. Using batch gradient descent (batch size is equal to the size of the training set) it took around 30 minutes and 4000 epochs to reach training loss of ~0.0002 on a GeForce GTX 1070 GPU.

To generate music, a sample unseen by the model during training with 40 timesteps is given as input and the single timestep predicted by the model is appended at $t$ = 41. Subsequently, timesteps 2–41 are given and the prediction is appended at $t$ = 42. This routine can continue in a loop for an arbitrary number of steps where the prediction at each timestep is always a function of the last 40 timesteps. This is a common approach for music generation systems that require a prompt before they can start the generation process.

For decoding the result into MIDI, first a desired soprano note must be specified as the opening. Then for each timestep the actual pitch of the soprano is obtained by adding the interval specified by the melody vector to the soprano pitch at the previous timestep. The rest of the pitches at each timestep are obtained by adding the intervals specified by the chord vector to the soprano from the same timestep.

## 6.3 Results and Discussion

Some generated outputs are available on SoundCloud for listening at https://soundcloud.com/hooman-rafraf/sets/deepdelta-samples. The results are musically convincing even though there are occasional dissonances that Bach would not write. Nevertheless, it is evident that the model learned contrapuntal voice-leading, tonal harmonic progression, cadences, fermatas, and musical phrasing quite well.

An interesting observation is the existence of frequent but smooth modulations in the generated music. This behavior was expected from the model from the start, and stems from the differential nature of the new data representation. A strong element in tonality is the musical key, which is largely specified by a certain set of pitches that the music is predominantly comprised of. Deviation from this set without performing a modulation has hard rules, especially in older genres such as Baroque music, and is typically temporary. After such deviations the music would normally reinforce the key by using a strong cadence on the tonic, otherwise the tonality would be lost. The model has no perception of a set of notes to adhere to. It only understands movements and intervals and as such, it will not necessarily return to the original key after each deviation. This results in compositions that contain noticeably more modulations than one would expect from chorale music. Be that as it may, since the model does learn the rules of counterpoint and voice-leading quite well, its transitions from one key to another are generally smooth and pleasant. By increasing the number of timesteps per training sample the problem of excessive modulations can be alleviated; but since our chorale dataset only contains short excerpts this experiment must be postponed for later endeavors.

Another issue to be addressed is the problem of register, i.e., how high or low the music is. The generated melody only contains a sequence of intervals, and the actual notes will be determined by the cumulative sums of this sequence. But the model has never been punished for creating a sequence whose cumulative sum exceeds too high or too low. Thus it is possible, especially in longer sequences, that the generated music spans an unacceptable range of notes. This problem can also to be mitigated by using longer samples during training and warrants future experimentation.

## 7  Conclusion and Future Work

I devised a new representation of music for training an LSTM-based AI model to compose music in the style of J. S. Bach. The idea is to learn the motion in melody and interval structure for chords instead of learning pitches. This approach will remove the need for data augmentation without loss of information, which significantly reduces the size of the training data. Another benefit offered by the new representation is the greater similarity between repeated patterns in music and their corresponding entries in the representation. These two advantages greatly facilitate the learning process. The model successfully learned to write tonal music in the style of Bach's chorales, respecting the rules of counterpoint and harmonic progressions as well as musical phrasing.

In the future I hope to improve the model by solving the problems of register and tonality: to make sure the music would stay in an acceptable range of pitches and does not modulate exceedingly to different keys. I also hope to use this encoding with a Transformer-based model in the future.